\documentclass[pra,aps,groupedaddress,twocolumn,floatfix,nofootinbib]{revtex4-2}
\usepackage{graphicx}
\usepackage{soul}
\usepackage{xcolor}

\newcommand{\beq}{\begin{equation}}
\newcommand{\eeq}{\end{equation}}
\newcommand{\beqa}{\begin{eqnarray}}
\newcommand{\eeqa}{\end{eqnarray}}

\def\opone{\leavevmode\hbox{\small1\normalsize\kern-.33em1}}

\begin{document}

\title{Elegance, Facts, and Scientific Truths}
\author{Nicolas Gisin \\
\it \small Group of Applied Physics, University of Geneva, 1211 Geneva 4,  Switzerland \\
Constructor University, Bremen, Germany}

\date{\small \today}
\begin{abstract}
I argue that scientific determinism is not supported by facts, but results from the elegance of the mathematical language physicists use, in particular from the so-called real numbers and their infinite series of digits. Classical physics can thus be interpreted in a deterministic or indeterministic way. However, using quantum physics, some experiments prove that nature is able to continually produce new information, hence support indeterminism in physics.
\end{abstract}
\maketitle

\section{Introduction}\label{intro}
Imagine a beautiful starry sky on a clear night with no light pollution. Humanity soon felt the need to tell the story of this marvellous scene. The idea that the heavens must be inhabited by perfect spheres came naturally. This image is elegant and resonates well with the fascination we feel in front of this spectacle. However, thanks to the first astronomical telescopes, Galileo saw mountains on the moon. He understood that the moon was not a perfect sphere. Similarly, Saturn is surrounded by rings. Eventually, everyone saw the mountains on the moon and realized that the stars were different from the ideal they had imagined. A few decades later, Newton understood that the terrestrial and celestial worlds obey the same laws, the theory of universal gravitation. The elegance of perfect spheres was replaced by the beauty of unifying worlds and the elegance of mathematical formulas:
\beq
F=m\cdot a=-\frac{G\cdot M\cdot m}{r^2}
\eeq
where $F$ and $G$ represent the force and constant of gravitation, respectively, $r$ the distance between masses $M$ and $m$, and $a$ the acceleration, i.e., the variation in velocity over time.

Mankind thus understands that it's the same force that makes apples fall, holds kangaroos to the ground and guides the moon in its orbit around the earth, as well as the earth around the sun. This unification is absolutely remarkable in its efficiency. Even today, rockets exploring outer space are guided by Newton's equation.

Newton's theory, which quickly took on the title of classical physics, seems to apply to everything from the simplest systems, such as the earth-moon pair and clocks, to the most complex systems, such as climate. Scientists concluded that this was the ideal of scientific explanation \cite{Anscombe71}. Intellectuals naturally deduced that the characteristics of classical physics, in particular Newton's equation, are characteristics of nature. According to this logic, since Newton's equation is deterministic, nature must itself be deterministic, as Laplace so aptly put it. So, since we can perfectly well calculate the dates and even the times of eclipses, both past and future, the climate and all other complex systems should be subject to the same scientific determinism: given the initial conditions, all the future and all the past are entirely determined. There is, of course, one “small” difference: for complex systems, there's no question of solving the equations, we can only calculate approximations or make statistical predictions. But since these complex systems are governed by the same equations, it's natural to assume that their evolution is equally deterministic. So, in classical physics, everything is determined, there is no creativity, no spontaneity. But is this really so, or are we blinded by the elegance of classical physics? In short, is scientific determinism a consequence of the facts, or of our desire for elegant explanations?

\section{Chaotic systems}
We've all heard of the butterfly effect, that tiny flap of a wing in Brazil which, as a result of an extremely complex chain of interactions between atmospheric molecules, ends up influencing the weather in Europe, eventually triggering a storm. This butterfly effect is a fine example of chaotic systems. Generally speaking, chaotic systems are characterized by their hyper-sensitivity to initial conditions: the slightest change in initial conditions can totally alter the system's evolution. This evolution is therefore extremely unstable. Let's illustrate this with an unrealistic model, simplified enough for everyone to understand the significance of this hyper-sensitivity to initial conditions. Let's imagine that the weather is reduced to a simple number between 0 and 1, such as the probability of rain. Let's denote this probability by the letter $x$. So, $x=0$ means there's no chance of rain, and $x=1$ means it's sure to rain. The value of $x$ changes from hour to hour. Let's assume, to illustrate a chaotic system, that the evolution of this system is such that the probability of rain at time $t+1$, denoted $x(t+1)$, can be obtained from the probability of rain at time $t$, denoted $x(t)$, simply by removing the first decimal place from $x(t)$ and shifting all the other decimal places one place to the left. By noting $d_1,d_2,d_3,...$ the decimal places of $x(t)$ we obtain the following simple evolution:
\beqa
x(t)\hspace{2mm} &=& 0,d_1d_2d_3......d_n... \nonumber\\
&&\hspace{5mm}\swarrow\swarrow...\swarrow \\
x(t+1)&=& 0,d_2d_3....d_n...  \nonumber
\eeqa
for example $x(t)=0.52314...$ becomes $x(t+1)=0.2314...$.
This evolution of the weather is not very realistic, but it illustrates the essential point of hyper-sensitivity to initial conditions. Indeed, at the initial instant $t_0$, the tenth decimal place, $d_{10}$, is of virtually no importance, and the thousandth decimal place, $d_{1000}$, even less so. However, ten hours later, $d_{10}$ has become the first decimal of $x(t_0+10)$, so it essentially describes the probability of rain: $d_{10}=9$ means that at time $t+10$ there is at least a 90\% chance of rain, and $d_{10}=0$ means that there is less than a 10\% chance of rain. Similarly, 6 weeks later, which corresponds to around 1000 hours, it's the thousandth decimal $d_{1000}$ of the initial condition that becomes decisive.

This example shows that decimals of the initial condition $x(t_0)$, which at the initial instant $t_0$ are of no importance, gain in importance over the course of the evolution until they practically determine the probability of rain. The same applies to more realistic complex systems, which is the very definition of a chaotic system.

So we can see that the future of any chaotic system is hidden in the insignificant decimals of the initial conditions (in our example, “initial condition” is singular because only one number describes our system). This raises the question of the relevance of insignificant decimals: do they have a physical reality, i.e. do they correspond to a physical reality, or are they merely the reflection of a mathematical language? In the former case, the evolution of chaotic systems is deterministic, in the latter case not necessarily. Since Galileo asserted that nature is written in mathematical language, and since in mathematics the so-called real numbers have an infinite number of decimal places, all well determined, the question no longer seems to arise. But many physicists have sensed that there is reason to doubt. For example, Max Born, one of the fathers of quantum mechanics, wrote \cite{Born}: {\it Statements like `a quantity x has a completely definite value' (expressed by a real number and represented by a point in the mathematical continuum) seem to me to have no physical meaning}.

It must be stressed here that these insignificant decimals are hidden from us, intrinsically inaccessible \cite{NGHiddenReals}. So it's not the facts that impose the reality of these decimals, but the mathematical language used in physics. Is it a leap of faith to assert their reality?

\section{What is a real number?}
A real number is a number with a decimal point. It begins with an integer, followed by a decimal point, then an infinite number of decimals. Sometimes these decimals stop, or more precisely, the sequence of decimals is continues with an infinite sequence of zeros, for example 1.500000... Generally, we truncate the 0s and simply write 1.5. Decimals of rational numbers always continue with an infinitely repeating sequence of decimals, such as 1/3=0.33333... But we all know of other real numbers, such as $\pi$, the ratio of the circumference to the diameter of a circle. In fact, how many real numbers do we know? A finite number, of course. And how many real numbers can we know? Knowing a number means we can give it a name. In a more technical but equivalent way, this “name” can be an algorithm that calculates the decimal of this number\footnote{More precisely, a series of approximations that get closer and closer to that number.}. An algorithm is a finite sequence of symbols, a sort of complicated but convenient name. So our question boils down to how many algorithms there are. The answer is relatively simple: we can number the names (and therefore the algorithms). Let's start with single-symbol names - there are as many as there are symbols. Then 2-symbol names, then 3-symbol names, and so on. There is therefore an infinite number of numbers that can be known, but a discrete infinity. Like integers, algorithms can be counted: 1, 2, 3, 4, ... The set of real numbers, on the other hand, is much larger, and is referred to as continuous infinity. This is not surprising, since real numbers can have an infinite number of decimals, and these decimals can have no structure at all. It's hard to conceive of an absence of structure, but this point is absolutely essential. If the decimals of a number have a structure, e.g. they repeat, then we can use this structure to name that number. So, the vast majority of real numbers have no structure. In short, there are calculable numbers and, infinitely more numerous, numbers without structure. The latter are called typical real numbers.

A first consequence is that the ``far'', ``insignificant'' decimals of typical real numbers are random. As complexity theorist Chaitin writes \cite{Chaitin1,Chaitin2}, the best way to conceive of a typical real number is to see it as the product of a true random generator. And it should be these real/random numbers that form the basis of scientific determinism? We're beginning to feel that something isn't quite as clear-cut as that. A second consequence of the structureless infinite series of decimals of typical real numbers, pointed out by the famous mathematician Borel \cite{Borel}, among others, is that a single real number can, in principle, contain in its infinite series of decimals the answers to all the (binary) questions that can be formulated in any human language. This is quite intuitive, since there is an infinite amount of information in each typical real number. Briefly, all you have to do is number all possible questions and encode the answers one after the other in the decimals of this “omniscient number”.

In short, there are two kinds of real numbers. Numbers that can be calculated using an algorithm, i.e. those with a name. And the infinitely more numerous numbers that have no names, that can't be calculated. Since the latter are infinitely more numerous than the former, we call them typical real numbers.

Let's try to imagine a typical real number. For simplicity's sake, let's imagine that it starts with a zero, followed by the decimal point and then the decimals. We can imagine these decimals, one after the other. With patience, you can imagine a lot of them. But it never ends, and there's no structure. So we have to imagine a sequence of decimals that goes on ad infinitum. If we imagine these decimals on a sheet of paper, we have to imagine a sheet so long that it overflows the earth, it overflows our galaxy, and it's never finished. That's how I intuitively imagine these numbers. But in fact, pure mathematicians conceive these numbers differently. For them, at least in principle, because I don't think any of them really do, you have to imagine all the decimals as given at once. Boom: an infinite amount of information given in a finite amount of time (in fact, zero time). 

This way of conceiving - and properly defining, it must be said - the real numbers of mathematicians is indisputably very elegant. But the question for the physicist is whether these numbers are physical, whether they represent a physical reality. In short, the challenge is to ask: does the determinism of classical physics result from facts, or from this mathematical language? Indeed, the numbers that mathematicians call real are physically random. A fact well summed up by the mnemonic slogan “Real numbers aren't really real” \cite{NGrealNb}.

\section{The importance of language, including mathematical language}
Mathematics is often written in the plural, because there are so many mathematical languages. However, at school and university, only one mathematical language is almost always taught: the classical one, which includes infinities and the law of the excluded middle. Many other mathematical languages reject infinities and the law of the excluded middle. These finitist mathematics are called constructivist mathematics. Here, it's not enough to postulate the existence of mathematical objects, such as infinite sequences of structureless decimals, but we limit ourselves to what can be constructed, at least in principle. But then, how to describe the continuum, how to represent a continuous set of points such as those constituting a straight line or a circle? This dilemma was resolved by the Dutch mathematician L.E.J. Brouwer over a century ago. Brouwer anticipated Chaitin's idea that the decimals of typical real numbers result from random processes (truly random, not pseudo-random like those of our computers). So, according to Brouwer's so-called intuitionist mathematics \cite{Brouwer1948,PosyBook,NPc,IntuitionismGisinSynthese}, decimals literally come one after another\footnote{More precisely, intuitionistic numbers progress over time, continuously gaining in precision. This is almost equivalent to new decimals gaining determined values one after the other, but not quite. A negligible point here \cite{BrassardDruckProblem}.}. Prior to their occurrence, their random production, they do not exist, at least not yet, their value is not yet determined. In a way, God or nature is continually playing dice, continually producing new information. So, at any given moment, there is only a finite quantity of information, a finite number of decimals. But this finite quantity progresses without limit. Classical mathematics and the continuum are found as a limit “at the end of time”. Everything can be calculated with intuitionistic mathematics, at least everything that is physical, everything that is measurable or observable at any given moment.

Brouwer's intuitionistic mathematics is not very popular today. Yet I believe that the majority of physicists, like myself, and computer scientists, necessarily, are doing intuitionistic mathematics without realizing it. Let's take the example of the climate, the importance of which we don't need to remind ourselves of today. How can scientists simulate the future climate? Obviously, they can't put real numbers as initial conditions into their computers, which are finite - gigantic, but finite. So they truncate the initial conditions, keeping only a finite number of decimal places. Then, in the course of their simulations, when the truncated decimals become important, they add random decimals \cite{PalmerStochClimateModel}. In short, they do intuitionistic mathematics.

A consequence of finitist mathematics, such as intuitionism, is that the law of the excluded middle is not valid: some propositions can be neither true nor false. For example, the proposition ``the decimal $d_n$ is 5'' can be neither true nor false if the value of that decimal has not yet been determined. This won't surprise anyone who believes in an open world, but it does upset many mathematicians who like to prove a theorem by simply showing that its negation is not tenable (the famous proofs by contradiction). However, for a physicist inclined to believe in an open world, and therefore in indeterminism, in order to prove that the weather will be fine a year from now, it's not enough to show that it's not certain that it won't rain; indeed, it may be that the weather in a year from now is not yet determined, and that different alternatives are still possible today.

To sum up, in intuitionistic mathematics, “nature” is capable of continuously producing new information, which is added - in finite quantities - to that which already exists. By contrast, in classical mathematics, instead of a “god” who continuously plays dice, god rolled all the dice at the big bang and encoded all the results in a monstrous real number defining the initial conditions of the universe. This is not to deny the elegance of Cantor's infinities, but simply to doubt that this is physics: our world is finite. Finite and continuously evolving.

It's clear, then, that determining whether Newton's classical mechanics is deterministic or not is not a scientific question; it depends on the meaning we attribute to the real numbers, and, ultimately, on the mathematical language we use. So far, the two alternatives - classical mathematics and classical deterministic physics, or intuitionistic mathematics and indeterministic physics - are both tenable options \cite{delsantogisin19}: it's a question of opinion, of a feeling of elegance, even of faith. To decide, we need the equivalent of Newton's astronomical telescope.

\section{The quantum physics ``telescope"}
No telescope or microscope will ever allow us to see the insignificant decimals directly - they're too far after the decimal point. However, quantum physics does allow us to “see” randomness, and thus to answer the question of determinism. Historically, quantum physics introduced the idea of randomness very early on. Who hasn't heard of Young's two-slit experiment, or Schrödinger's cat superposed between life and death? But the Young's slit experiment can be explained without any chance at all (it's de Broglie's pilot wave idea, developed by Bohm), and no cat has ever been seen in such a state, half a finger away from death. Quantum physics has much better to offer.

Imagine two physicists at a distance from each other \cite{Qchance}. In theory, this distance could be astronomical, but in practice it's limited to a few kilometers or tens of kilometers. The important thing is that they can't communicate during the experiment. This impossibility can be guaranteed by running the experiment so fast that communication would require a speed greater than the speed of light. But the absence of communication can also be guaranteed by physical constraints, such as a huge wall or field of bombs separating our two physicists. In practice, of course, we're content with common sense, although some experiments have gone so far as to impose that any communication would have to travel at least 100,000 times the speed of light \cite{BancalHiddenInfluence}. In short, let's imagine our two physicists at a distance with no means of communication. It's a hypothesis, an extremely natural hypothesis, but since our aim is to rigorously prove the existence of chance - and therefore of indeterminism - we need a mathematical theorem, and all theorems are based on hypotheses. So we'll assume that space does exist, and that therefore physicists who are far enough apart cannot communicate at arbitrarily high speeds. We need a second assumption. Our physicists must be able to choose “freely” which measurements they will carry out, each one a measurement. We leave each of them only the choice between two possible measurements. In fact, it's not even necessary for their choices to be of their own free will; it's enough for their choices to be independent of each other and of the system on which they're going to carry out their measurements. For example, the experiment has been prepared by a third party, and our two physicists' choices dictated by two old digitized films, one by Charlie Chaplin, the other by Louis Lumière \cite{PironioInputs}. In such a case, it's practically self-evident that the choices of measurements are independent of each other and of the way in which the experiment has been prepared.

The above description may seem extremely cumbersome, even pedantic. However, it is necessary because this quantum ``telescope"-experiment has actually been carried out \cite{Clauser72,Aspect82,Gisin98,Weihs98}, it has even resulted in the 2022 Nobel Prize in Physics. What's more, this quantum physics experiment leads to the rigorous conclusion that our two physicists' measurement results are necessarily random. This conclusion follows logically from the two hypotheses summarized above, as well as from observation of the statistics of the results obtained by the two physicists\footnote{Today many experiments have obtained statistics close enough to theoretical predictions, what physicists call a violation of Bell's inequalities, for this conclusion to be inevitable.} \cite{Acin2016}. It's impossible to overestimate the importance of this rigorous conclusion, which is unfortunately still all too often ignored. 

Thus, the “quantum telescope” makes it possible to see “mountains” on the ideal of scientific determinism: this determinism is not smooth, in fact it is untenable. Stars are not perfect spheres, and nature - even if limited to physics - is not deterministic.

\section{Conclusion}
Classical physics can be interpreted in a deterministic or indeterministic way, depending on the status we attribute to real numbers and which mathematical language we favour. Ultimately, it's a question of elegance. The elegance of determinism and real mathematical numbers, versus the elegance of an open world and intuitionistic mathematics.

On the other hand, to maintain determinism in quantum physics, one must abandon at least one of the two hypotheses formulated above, i.e. either abandon the notion of distance - or equivalently allow for infinite speeds - or abandon the idea that certain phenomena - an old film and a source of quantum particles - are independent of each other. Some of my colleagues are still working on abandoning one or the other of the above hypotheses. But frankly, determinism is dead: we've seen mountains, irregularities on this beautiful old idea of determinism. Nature is capable of creative acts, of continually creating new information, of producing events that were not necessary. At least, that's clearly my preferred choice. To abandon either of the above hypotheses is repugnant to me. Elegance forces quantum chance on us, although it's hard to imagine \cite{Qchance}.

In conclusion, as we have seen from the example of determinism in physics, it seems that scientific truths are not only based on facts, but also on our violent desire for elegance.

\small
\section*{Acknowledgment} This work has been supported by the Swiss National Science Foundation within the NCCR SwissMap. \\

\end{document}